# High-pressure synthesis of superconducting $Sn_3S_4$ using diamond anvil cell with boron-doped diamond heater

*Ryo Matsumoto[a], Kensei Terashima[b], Satoshi Nakano[c], Kazuki Nakamura[b,e], Sayaka Yamamoto[b,e], Takafumi D. Yamamoto[b], Takahiro Ishikawa[d], Shintaro Adachi[f], Tetsuo Irifune[g], Motoharu Imai[c], and Yoshihiko Takano[b,e]

*CA: MATSUMOTO.Ryo@nims.go.jp

[a]International Center for Young Scientists (ICYS),
National Institute for Materials Science, Tsukuba, Ibaraki 305-0047, Japan
[b]International Center for Materials Nanoarchitectonics (MANA),
National Institute for Materials Science, Tsukuba, Ibaraki 305-0047, Japan
[c]Research Center for Functional Materials,
National Institute for Materials Science, Tsukuba, Ibaraki 305- 0044, Japan
[d]Elements Strategy Initiative Center for Magnetic Materials (ESICMM),
National Institute for Materials Science, Tsukuba, Ibaraki 305-0047, Japan
[e]University of Tsukuba, Ibaraki 305-8577, Japan
[f]Nagamori Institute of Actuators, Kyoto University of Advanced Science, Ukyo-ku, Kyoto 615-8577, Japan
[g]Geodynamics Research Center (GRC), Ehime University, Matsuyama, Ehime 790-8577, Japan
[f]Research Center for Functional Materials,

**Abstract**

High-pressure techniques open exploration of functional materials in broad research fields. An established diamond anvil cell with a boron-doped diamond heater and transport measurement terminals has performed the high-pressure synthesis of a cubic $Sn_3S_4$ superconductor. X-ray diffraction and Raman spectroscopy reveal that the $Sn_3S_4$ phase is stable in the pressure range of $P>5$ GPa in a decompression process. Transport measurement terminals in the diamond anvil cell detect a metallic nature and superconductivity in the synthesized $Sn_3S_4$ with a maximum onset transition temperature ($T_c^{onset}$) of 13.3 K at 5.6 GPa. The observed pressure-$T_c$ relationship is consistent with that from the first-principles calculation. The observation of superconductivity in $Sn_3S_4$ opens further materials exploration under high temperature and pressure conditions.

# 1. Introduction

Modification of crystal and electronic structures by applying external fields, such as electric field and pressure, reveals the hidden functionalities in materials. Binary tin chalcogenides $Sn_xCh_y$ ($Ch$ = S,Se) exhibit various functionalities under the external fields, for example, enhancement of thermoelectric performance[1,2], appearance of topological Dirac line node[3], and superconductivity[4-6]. Among the $Sn_xCh_y$, SnS and SnSe ($x = y = 1$) show pressure-induced superconductivity with a transition temperature ($T_c$) of 5.8 K at 47.8 GPa[7] and 3.2 K at 39 GPa[8]. $SnCh_2$ has been investigated as a promising candidate for 2D characteristic materials[9,10], and superconductivity is induced in $SnSe_2$ by applying an electric field[4,5]. The emergence of superconductivity in $SnS_2$ by Li-intercalation has also been predicted based on calculations[11]. Although the $Sn_2Ch_3$ compounds are recently focused on phase-change memory[12] and solar cell material[13], there is no report on the emergence of superconductivity under external fields.

A pressure-induced phase stabilization and an emergence of superconductivity have been predicted in a cubic $Sn_3Ch_4$ structure with $I\text{-}43d$ space group from a first-principles evolutionary crystal structure search[14,15]. According to a formation energy analysis, $Sn_3Se_4$ is energetically stable above 10 GPa[14]. The calculations of electronic band structure and phonon dispersions suggest that $Sn_3S_4$ has a metallic nature in transport property, and it would exhibit superconductivity based on the phonon-mediated Bardeen–Cooper–Schrieffer (BCS) mechanism with $T_c$ of 3.3-4.7 K at 10 GPa. Recently, the cubic $Sn_3Se_4$ has been synthesized at 16.4 GPa and 1225 K via laser heating in a diamond anvil cell (DAC). However, the physical properties of $Sn_3Se_4$, including superconductivity, are not measured because it decomposes at ambient pressure[14]. The in situ measurement of electrical resistivity or magnetization under high pressure is required to observe superconductivity. On the other hand, cubic $Sn_3S_4$ is also predicted to be stable at 15 GPa with metallic nature and superconductivity with $T_c$ of 12 K[15]. To the best of our knowledge, no reports have been made on the experimental synthesis of the cubic $Sn_3S_4$. By analogy with the case of $Sn_3Se_4$, the $Sn_3S_4$ phase is also considered to be stable only at high pressure and unquenchable as a metastable phase at ambient pressure. The high-pressure synthesis and in situ electrical measurement should be performed simultaneously to reveal the theoretically predicted superconductivity in the $Sn_3Ch_4$ system.

A combination technique of high-pressure synthesis and in situ transport measurement has been recently established using the DAC with a heater, thermometer, and measurement terminals comprised of the boron-doped diamond (BDD) epitaxial film[16]. There, the pressure range was limited to a relatively lower value because a box-type diamond anvil and a soft backup plate of pyrophyllite are used as DAC components. To achieve higher pressure, a replacement of the box-type anvil to a culet-type anvil. Moreover, the backup plate made from good thermal insulating material with higher toughness is necessary. In the current study, we fabricate the BDD heater, thermometer, and measurement terminals on the culet-type diamond anvil with the backup plate of $ZrO_2$ that enabled us to perform the high-pressure synthesis above 30 GPa. We have successfully synthesized the cubic $Sn_3S_4$ using this improved DAC with the BDD heater and observed the emergence of superconductivity under pressure ranging from 24.6 to 5.6 GPa. It is also found that the $T_c$ of $Sn_3S_4$ increases with decreasing pressure, and the onset of $T_c$ ($T_c^{onset}$) and temperature at zero resistivity ($T_c^{zero}$) take the maximum value of 13.3 and 8.5 K at 5.6 GPa.

## 2. Experimental procedures
### 2.1. Preparation of diamond anvil cell

The patterns for BDD deposition were drawn by an electron beam lithography and a microwave plasma-assisted chemical vapor deposition (MPCVD). The boron concentrations of the heater and the terminals were tuned above an order of $10^{21}$ cm$^{-3}$ to obtain a metallic diamond[17-19]. The transport property of the BDD thermometer was adjusted to show semiconducting nature by reducing the source gas of boron in the MPCVD process. Although the fabrication process of the BDD components is applicable to any kinds of diamond, a nano-polycrystalline diamond is more suitable for the generation of high temperature and pressure owing to superior hardness and thermal insulation[20,21]. The details of the fabrication process and the properties of the BDD components are described in elsewhere[22,23]. The diamond anvils were fixed to the $ZrO_2$ backup plates using Ag paste. The typical diameters of the anvil culet and gasket hole were 300 μm and 200 μm, respectively. A sample space was composed of a SUS316 stainless steel or Re gasket, cubic boron nitride (cBN) pressure-transmitting medium, and pressure sensor of ruby powder. The generated pressure was determined using the ruby fluorescence method[24] and the Raman shift of diamond[25] at room temperature. The error of the pressure determination was estimated by referring to the reports.

### 2.2. Sample characterization

The crystal structure of the sample in the DAC was investigated through in situ XRD measurements using synchrotron radiation performed at AR-NE1A beamline of the Photon Factory (PF) at the High Energy Accelerator Research Organization (KEK). The X-ray beam was monochromatized to an energy of 30 keV ($\lambda = 0.4180$ Å) and introduced to the sample in the DAC through a collimator with 50 μm diameter. The wavelength was calibrated by using a $CeO_2$ standard. The diffraction patterns were obtained by the imaging plate type detector installed in the downstream side of the X-ray with an exposure time of 60 min for each pressure. The XRD patterns were integrated into a one-dimensional profile using the IPAnalyzer[26]. The XRD patterns of the sample were taken from 41.1 GPa to ambient pressure at room temperature. For comparison, the XRD patterns of starting materials under compression and decompression processes without heating were also acquired between ambient pressure and 39.2 GPa. The lattice constant of the products under various pressures was evaluated by fitting the identified diffraction peaks to the calculation pattern using PDIndexer[26] after the background subtraction through QualX2[27]. The bulk modulus of the obtained phase was calculated by fitting the third-order Birch-Murnaghan equation of state on the pressure dependence of unit cell volume using the fitting program EOSFIT5.2[28]. The diffraction patterns simulation was performed using VESTA[29]. The vibrational modes of the sample before and after the heating were examined through in situ Raman spectroscopy measurements using an inVia Raman microscope (RENISHAW) with a laser wavelength of 532 nm. The reproducibility of the measured wave number was within ±0.1 cm$^{-1}$. The electrical transport measurement was conducted in the temperature range of 2-300 K using a physical property measurement system with a 7 T superconducting magnet (PPMS/Quantum Design).

The presented data of resistance measurement during the synthesis and in situ XRD measurements after the synthesis were acquired in the same experimental run (labeled by #1). The pressure profile of XRD for the starting materials without heating was measured in run #2. The Raman spectroscopy and evaluation of the superconducting properties were also performed in run #3. The high-pressure synthesis was reproducibly performed in run #4 using the same configuration of DAC with #3.

## 2.3. Computational details

The electronic structures of $Sn_3S_4$ under various pressures were calculated using Quantum ESPRESSO (QE)[30-32]. The generalized gradient approximation (GGA) of Perdew-Burke-Ernzerhof (PBE)[33] was used to describe the exchange-correlation function with the ultra-soft pseudopotential through the Rappe Rabe Kaxiras Joannopoulos method[34]. Before the electronic structure and phonon calculations, the stable atomic positions and the lattice constant were calculated under each pressure. A $6 \times 6 \times 6$ $k$-grid was employed for the $k$-point sampling in the first Brillouin zone, and the kinetic energy cutoffs for the expansion of the electronic wave function were set to 80 Ry. A $k$-point mesh of $12 \times 12 \times 12$ was used to calculate the density of states (DOS). Electron-phonon coupling and average logarithmic phonon frequency under pressures were calculated using QE[30-32], and the $T_c$ values were calculated using the Allen-Dynes modified McMillan formula[35,36], where a q-mesh of $6 \times 6 \times 6$ was considered. The frequencies of Raman active modes were calculated by the finite displacement method as implemented in the Phonopy code[37] and QE with a $2 \times 2 \times 2$ supercell, a lattice constant of 7.702 Å, and a $k$-point mesh of $3 \times 3 \times 3$.

## 3. Results and discussion
### 3.1 High-pressure synthesis

Cubic $Sn_3S_4$ is synthesized under high temperature and high pressure using the diamond anvil with BDD heater, thermometer, and measurement terminals, as shown in Fig. 1 (a). The heater and thermometer are designed on the slope part of the diamond anvil. The DAC is prepared using the fabricated anvil for the high-pressure synthesis of $Sn_3S_4$, as shown in Fig. 1 (b). As a sample, the well-ground mixture of SnS and SnS$_2$ with a stoichiometric composition of Sn$_3$S$_4$ is placed on the BDD terminals on the top anvil. After the compression above 30 GPa, the sample is heated up to about 800 K using the BDD heater and then cooled to 300 K with a continuous measuring of the sample resistance in a glass tube under N$_2$ gas flow to avoid diamond oxidation. The temperature of the sample was measured using both the BDD thermometer and a radiation thermometer through an optical fiber cable.

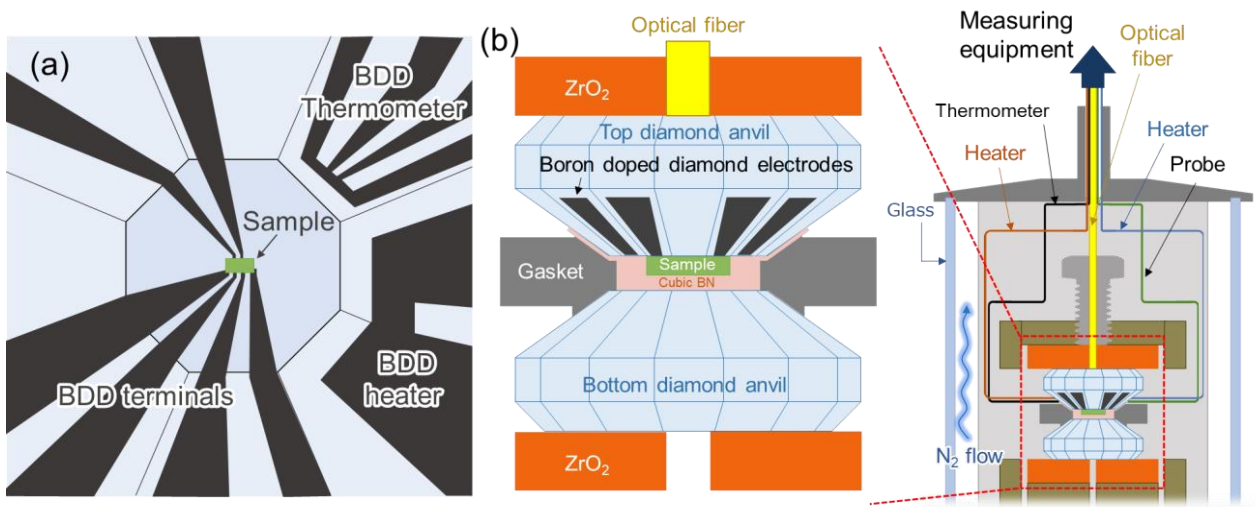

**Figure 1. (a) Schematic image of the fabricated diamond anvil with boron-doped diamond (BDD) components of the measurement terminals, heater, and thermometer. (b) Configuration of the diamond anvil cell (DAC) and the sample synthesis environment.**

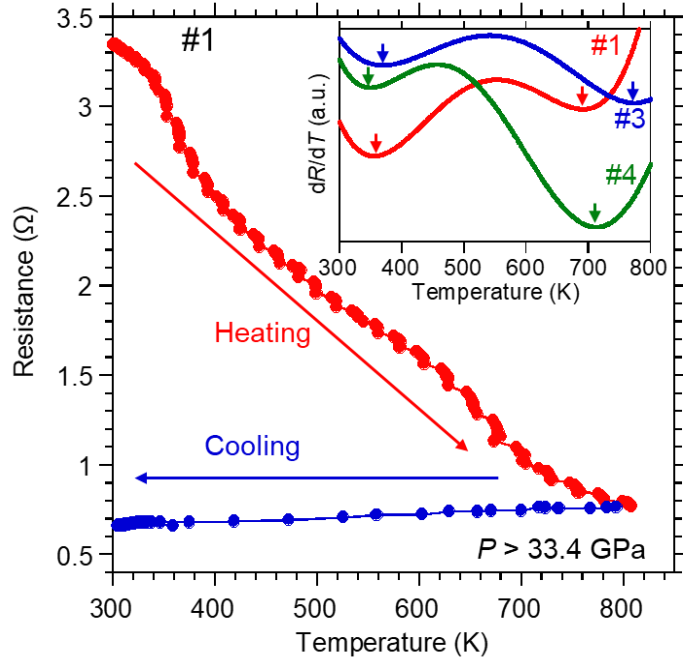

**Figure 2. Resistance of the sample in the DAC as a function of the temperature during the synthesis under pressure above 33.4 GPa. The inset is the temperature derivative of resistance on the heating process for various experimental runs with different pressures of 33.4 (#1), 29.5 (#3), and 35.7 GPa (#4).**

Figure 2 shows a measured resistance of the mixture of the starting materials during the high-pressure synthesis above 33.4 GPa. The sample temperature increases up to about 800 K when the input power to the BDD heater is about 19 W. The sample exhibits non-metallic behavior in the heating process, in which the slope of the resistance changes twice at around 350 K and 700 K. The temperature derivative of resistance in the heating process for various experimental runs with different pressures of 33.4 (#1), 29.5 (#3), and 35.7 GPa (#4) are shown in the inset of Fig. 2. The anomalies are reproduced in all the runs and tend to appear at lower temperatures under high pressures (#1 and #4) compared with the lower pressure (#3). The input power of the BDD heater is gradually reduced after confirming the second anomaly without waiting time at maximum temperature. The sample shows metallic behavior in the cooling process, and the resistance at 300 K is drastically decreased by 20% after the synthesis. The anomalies and irreversibility of resistance observed here imply that a solid-state reaction in the starting materials has occurred under high pressure and high temperature. The observed metallic property is different from both starting materials of SnS and SnS$_2$ with the semiconducting band structures[38,39]. The pressure is increased from 33.4 GPa to 41.1 GPa during the synthesis.

**3.2. Structural analysis**

The in situ X-ray diffraction (XRD) measurements were performed to investigate the crystal structure of the synthesized product. Debye-Scherrer diffraction rings are observed in all acquisitions with a maximum 2θ angle of 16º, as shown in the representative data at 41.1 GPa in Fig. S1. Figure 3 (a) shows a comparison of the XRD patterns before and after the synthesis. The XRD pattern of the heated sample inside the DAC is completely different, and the peaks labeled by the triangle and square markers appear. The highest intensity peak at ~ 11° and the peak marked by a circle are due to the cubic boron nitride used as a pressure-transmitting medium. The triangle-labeled peaks can be indexed as the theoretically predicted cubic $Sn_3S_4$ ($I$-$43d$)[15], of which the simulated peaks are shown in Fig. 3 (a). Figure 3 (b) shows the XRD patterns of the sample after the synthesis at various pressures in the decompression process. The peaks of $Sn_3S_4$ gradually shift to the lower $2\theta$ side as the pressure decreases, indicating a lattice expansion. Under the decompression process, the $Sn_3S_4$ phase is stable in a wide pressure range (6.2 GPa<$P$<41.1 GPa). By further releasing the pressure to ambient pressure, the $Sn_3S_4$ peaks disappear, suggesting a decomposition into the starting materials. The peaks with the square marker are in agreement with those of the theoretically predicted high-pressure phase of SnS ($Pm$-$3m$)[15]. According to the calculation of formation enthalpy[15], the $Sn_3S_4$ is more stable than the high-pressure phase of SnS until 60 GPa. Therefore, we consider that the high-pressure phase SnS observed in XRD is formed from the unreacted residue of the starting material. As focused on a representative peak of SnS with the arrows in Fig. 3 (b), the structural transition is occurred in SnS from the high-pressure cubic phase to the lower pressure phase below 27.0 GPa.

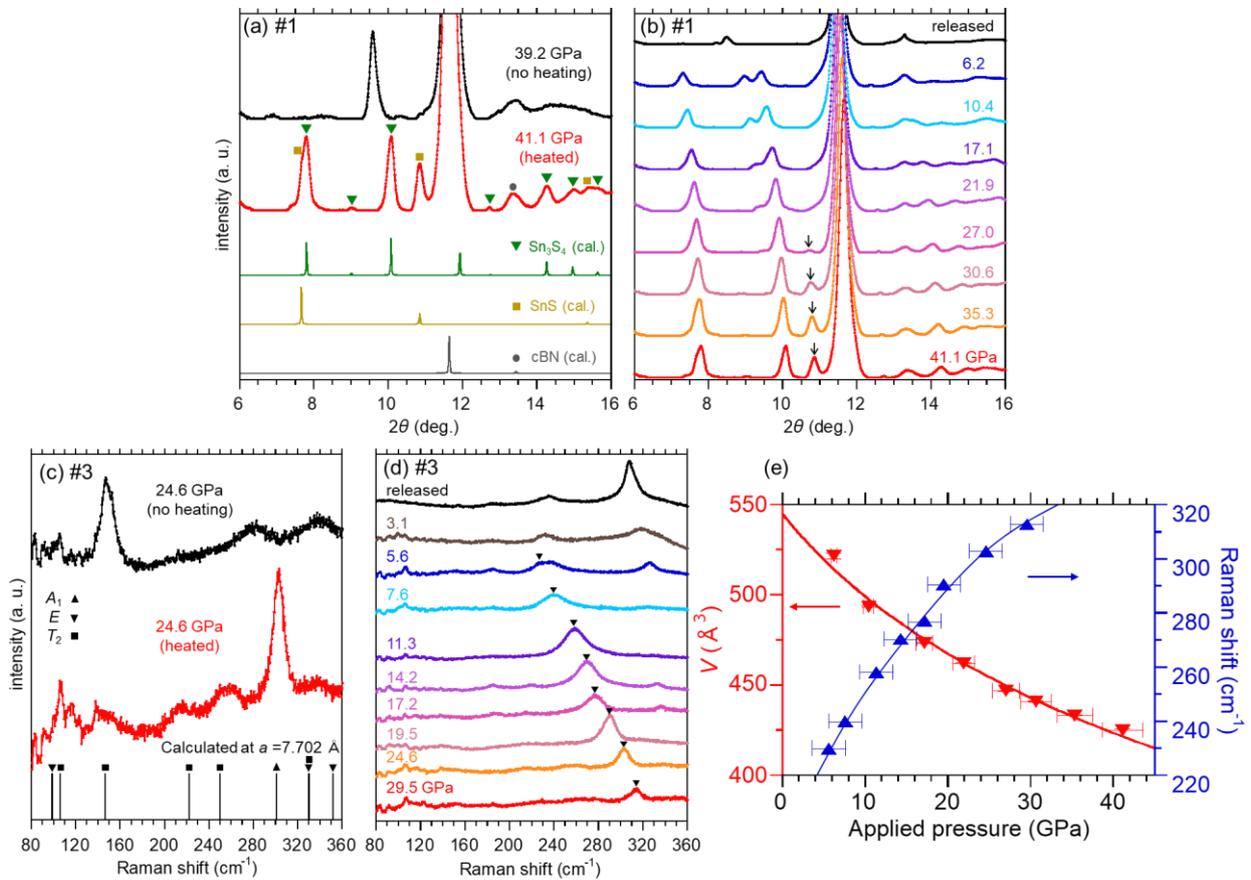

**Figure 3.** (a) Comparison of the X-ray diffraction (XRD) patterns before and after the synthesis. The simulation patterns of cubic $Sn_3S_4$, cubic SnS, and cubic BN are shown at the bottom. (b) XRD patterns of the sample after the synthesis at various pressures in the decompression process. (c) Comparison of the Raman spectrum of the sample at 24.6 GPa before and after the synthesis. Calculated Raman active modes of $Sn_3S_4$ at the lattice constant of $a = 7.702$ Å are shown in the bottom. (d) Raman spectra of the synthesized sample under the process of decreasing pressure from 29.5 GPa to ambient pressure. (e) Applied pressure dependence of a unit cell volume $V$ and the position of Raman peak of $A_1$ mode in $Sn_3S_4$. The fitted curve for the $V$ is the Birch-Murnaghan equation of state.

The Raman spectroscopy was conducted for the further characterization of the products. Figure 3 (c) shows a comparison of the Raman spectrum of the sample at 24.6 GPa before and after the synthesis with the positions of calculated Raman active modes of $Sn_3S_4$. The Raman peaks of the starting materials vanish and some peaks newly appear via the high-pressure synthesis. The symmetry analysis shows that four kinds of Raman-active modes ($A_1$, $E$, and $T_2$) exist at the gamma point. The peak located at about 300 cm$^{-1}$ with triangle marker is clearly visible in the analysis region (80-400 cm$^{-1}$), which is identified as $A_1$ symmetry. The other observed peaks can also be assigned by the peaks derived from $E$ and $T_2$ modes, although the intensities are tiny. Figure 3 (d) displays the Raman spectra of the synthesized sample under the decompression process from 29.5 GPa to ambient pressure. The Raman peaks of cubic $Sn_3S_4$ disappear below 3.1 GPa, and the spectrum corresponding to the starting powder is recovered at ambient pressure. The decomposition feature is consistent with the result of the XRD analysis. Here, there is no Raman peak of cubic boron nitride in the analyzed region[40]. The pressure profiles of XRD patterns and Raman spectra of the starting material are presented in Fig. S2.

The results of the XRD and Raman analysis are summarized in Fig. 3 (e). The left axis shows the unit cell volume of the $Sn_3S_4$ phase with the curve fitted to the Birch-Murnaghan equation of state, assuming that the first pressure derivative of the bulk modulus $B_0'$ is 4.0. The bulk modulus $B_0$ and the unit cell volume $V_0$ at 0 GPa are evaluated from the fitting to be 97(6) GPa and 545(5) Å$^3$, respectively. The unit cell volume tends to be smaller than theoretical prediction [15] due to an overestimation feature in the used calculation method[41]. The right axis shows the position of the Raman peak corresponding to $A_1$ mode. The Raman peak of $A_1$ mode shifts to the lower frequency side with decreasing pressure, indicating a phonon-softening of cubic $Sn_3S_4$.

### 3.3. Superconductivity

Figure 4 (a) shows the temperature dependence of resistance for $Sn_3S_4$ under the decompression process after the high-pressure synthesis. The sample exhibits metallic behavior under all pressure regions. As shown in Fig. 4 (b), the sharp drop of resistance is observed at low temperatures in the wide pressure range from 24.6 to 5.6 GPa, corresponding to the superconducting transition. The resistance is increased just before the onset of the superconducting transition, which is a known feature for the inhomogeneity of the superconducting region[42]. The theoretical prediction[15] for Sn-S compounds suggests the following two candidates for superconductor: cubic $Sn_3S_4$ and the high-pressure phase of SnS. According to the XRD analysis, the high-pressure phase of SnS is recovered to a non-superconducting lower pressure phase below 27 GPa. In contrast, the cubic $Sn_3S_4$ phase can exist down to 5.6 GPa, as is observed in the Raman spectra. Therefore, we conclude that the observed superconducting transition is attributed to the cubic $Sn_3S_4$.

Figure 4 (c) shows the resistance of $Sn_3S_4$ as a function of the temperature at 5.6 GPa under magnetic fields up to 7 T. The $T_c$ of $Sn_3S_4$ gradually decreases with increasing magnetic fields, which is further evidence of superconductivity. The inset of Fig. 4 (c) shows the temperature dependence of the upper critical field $B_{c2}$. The $T_c$ is determined as the temperature at which the 50% drop of the normal resistance is observed[43]. The $B_{c2}(0)$ is estimated to be 10.4 T under 5.6 GPa from the parabolic fitting. From the Ginzburg-Landau (GL) formula $B_{c2}(0) = \Phi_0/2\pi\xi(0)^2$, where the $\Phi_0$ is a fluxoid, the coherence length at zero temperature $\xi(0)$ is estimated to be 5.9 nm. Note that the resistance drop starts at around 13 K, as shown in expanded plots of Fig. 4 (d). Such a high $T_c^{onset}$ is a signature of a higher $T_c$ portion in the sample. Considering that the $T_c$ increases with decreasing pressure, pressure distribution may have caused local low-pressure regions that might have led the formation of the filamentary high-$T_c$ state.

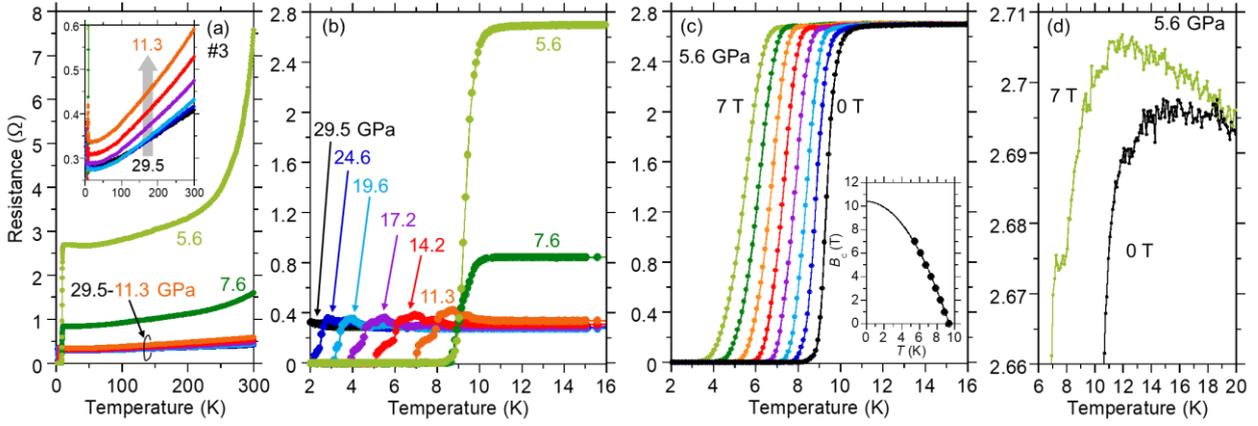

**Figure 4.** (a) Temperature dependence of resistance in $Sn_3S_4$ with decreasing pressure from 29.5 to 5.6 GPa. The magnified plots for the low-pressure region are shown in the inset. (b) The expanded plots of the measured resistance at the low-temperature region. (c) Temperature dependence of resistance at 5.6 GPa under various magnetic fields. The inset shows a temperature dependence of the upper critical field $B_{c2}$. (d) Expanded plots of the temperature dependence of resistance at around $T_c^{onset}$.

The electronic and phonon structure under various pressures were evaluated through the first-principles calculation to understand the decompression-driven enhancement of $T_c$ observed in the $Sn_3S_4$. Here, our pressure calculation also tends to overestimate, as seen in the previous report[15]. The calculated pressures have been replaced by the experimentally observed values at the same lattice constant. Figure 5 (a) shows a calculated electronic density of state (DOS) under pressures from 7.9 GPa to 43.3 GPa. The DOS under all the pressures are metallic, which is consistent with the calculated band structure in the previous report[15]. The decreasing pressure induces a band narrowing, and consequently, the $N(0)$, defined by DOS at Fermi energy ($E_F$), increases under decompression. Although the enhancement of $N(0)$ generally reduces the electrical resistance, it tends to increase when the pressure decreases in the experiment, as shown in Fig. 4 (a). The degradation of conductivity is possibly due to an extrinsic reason such as weakening of grain boundary or a partial decomposition. The decompression-induced enhancement of $N(0)$ increases the $\lambda$, as shown in Fig. 5 (b). As a result, the $\lambda$ can be drastically increased with decreasing pressure, as seen in the left axis of Fig. 5 (c). In contrast, as shown in the right axis of Fig. 5 (c), the $\omega_{log}$ is reduced by decreasing pressure because of a phonon softening, which is consistent with the experimental results of the Raman spectroscopy. The $T_c$ of the phonon-mediated BCS superconductor can be described by the following Allen-Dynes

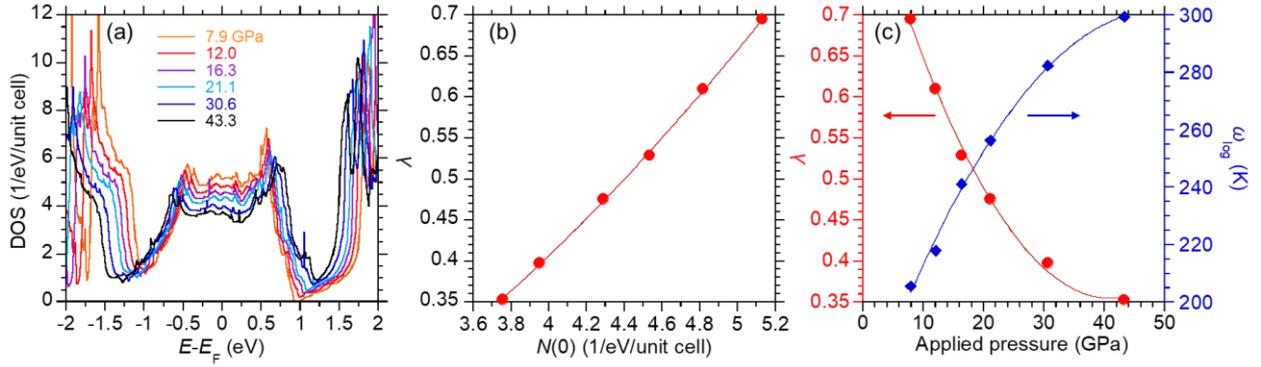

**Figure 5. (a) Electronic DOS under the pressures from 7.9 to 43.3 GPa. (b) Relationship between the DOS at Fermi energy ($N(0)$) and the electron-phonon coupling constant ($\lambda$). (c) Applied pressure dependence of $\lambda$ and average logarithmic phonon frequency ($\omega_{\log}$).**

modified McMillan formula[35,36],

$$T_c = \frac{\omega_{\log}}{1.2} \exp\left[-\frac{1.04(1+\lambda)}{\lambda - \mu^*(1+0.62\lambda)}\right], \qquad (1)$$

where $\lambda$ is the electron-phonon coupling constant, and $\omega_{\log}$ is the average logarithmic phonon frequency. In this formalism, $\mu^*$ is the Coulomb repulsion parameter with typical values of 0.1. In most cases, the contribution of $\lambda$ is higher than that of $\omega_{\log}$ on the change of $T_c$ in BCS superconductors, as seen in MgB$_2$[44]. The calculated $T_c$ of Sn$_3$S$_4$ monotonically increases with decreasing pressure based on the McMillan formula, reflecting the enhancement of $\lambda$.

Figure 6 (a) shows the pressure dependence of $T_c^{\text{onset}}$, $T_c^{\text{zero}}$, and calculated $T_c$ ($T_c^{\text{calc}}$) on Sn$_3$S$_4$. As a result of the enhancement of $\lambda$ and the reduction of $\omega_{\log}$, the $T_c^{\text{calc}}$ increases with decreasing pressure, which shows a good agreement with the experimentally observed $T_c$s. The maximum $T_c^{\text{zero}}$ and $T_c^{\text{onset}}$ are 8.5 K and 13.3 K at about 5 GPa. In the decompression process, the cubic Sn$_3$S$_4$ phase synthesized above 25 GPa is stable in the pressure range of $P > 5$ GPa. The correspondence between the experimental and calculated $T_c$s suggests that $Sn_3S_4$ is a conventional BCS superconductor, as predicted by the previous report[15]. A lighter element substitution for the Sn site, for example, Ge or Si, is expected to enhance $T_c$ based on the BCS perspective.

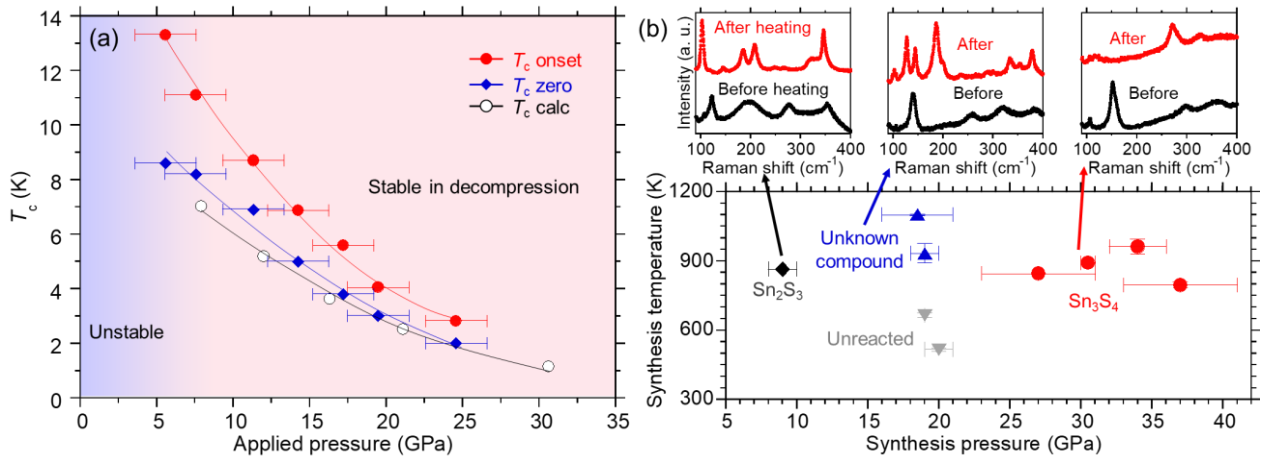

**Figure 6. (a) Pressure dependence of $T_c^{\text{onset}}$, $T_c^{\text{zero}}$, and $T_c^{\text{calc}}$ in Sn$_3$S$_4$. (b) Synthesized products under different pressure and temperature and the representative Raman spectrum on each phase.**

Various synthesis conditions of temperature and pressure were examined to investigate a phase diagram of the Sn-S binary system. Figure 6 (b) shows the synthesized products under different synthesis conditions and the representative Raman spectrum on each phase. The $Sn_3S_4$ phase is synthesized above 25 GPa with good reproducibility. On the other hand, the appeared peaks below 10 GPa correspond to the $Sn_2S_3$ phase (See Fig. S3)[45]. The observed spectrum at around the synthesis pressure of 20 GPa can not be explained by the possible candidate materials of SnS, $SnS_2$, $Sn_2S_3$, and $Sn_3S_4$. The investigation of crystal structure and physical properties in the unknown compound are expected in future works.

## 4. Conclusion

In this study, the theoretically predicted superconductor cubic $Sn_3S_4$ has been synthesized using the DAC with a BDD heater under high-temperature and high-pressure conditions. The synthesized $Sn_3S_4$ phase at above 25 GPa is stable in the pressure range of $P>5$ GPa in the decompression process. The observed $T_c$ shows the positive pressure effects against decreasing pressure, and the maximum $T_c^{zero}$ and $T_c^{onset}$ are 8.5 K and 13.3 K at about 5 GPa. The decompression-induced enhancement of $T_c$ can be explained by the first-principles calculation and it originates from the increasing $\lambda$. The DAC with BDD heater and measurement terminals are suitable to complete the phase diagram for the synthesis of materials. Further investigation of synthesis for $Sn_xCh_y$ binary compounds under high pressure is expected using the established techniques.

**Acknowledgment**
This work was partly supported by JST CREST (Grant No. JPMJCR16Q6), JST-Mirai Program (Grant No. JPMJMI17A2), JSPS KAKENHI (Grant No. 19H02177, 20H05644, and 20K22420). The fabrication of the diamond electrodes was partly supported by NIMS Nanofabrication Platform in Nanotechnology Platform Project sponsored by the Ministry of Education, Culture, Sports, Science and Technology (MEXT), Japan. The synchrotron X-ray experiments were performed at AR-NE1A (KEK-PF) under the approval of Proposal No. 2020P008 with support from Dr. Y. Shibazaki (KEK). The high-pressure experiments were supported by the Visiting Researcher's Program of the GRC. The nano-polycrystalline diamond was synthesized by Toru Shinmei of the GRC. The calculation of the Raman peak was supported by Dr. S. Tsuda (NIMS). RM would like to acknowledge the ICYS Research Fellowship, NIMS, Japan.

**Supporting Information**
Typical Debye-Scherrer diffraction ring in high-pressure XRD analysis, XRD patterns and Raman spectra from the starting materials, and Raman spectra in $Sn_2S_3$.

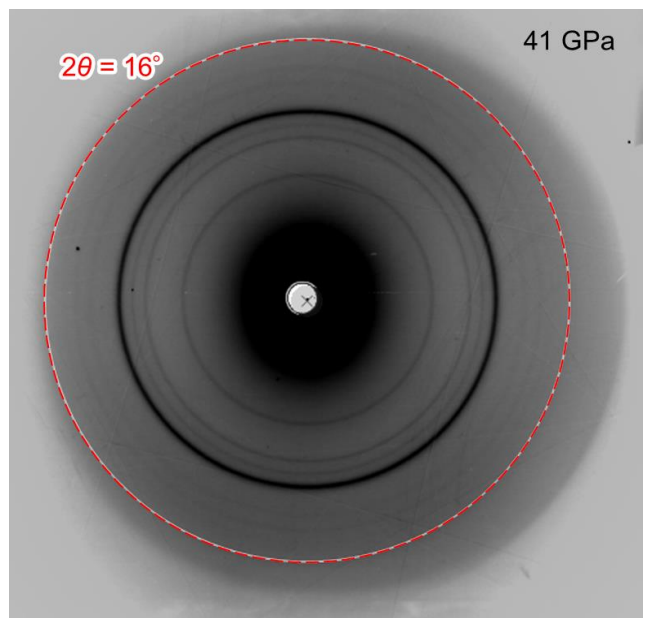

**Figure S1.** Representative Debye-Scherrer diffraction ring (at 41 GPa). The dashed circle is a guide for a maximum $2\theta$ angle of 16º.

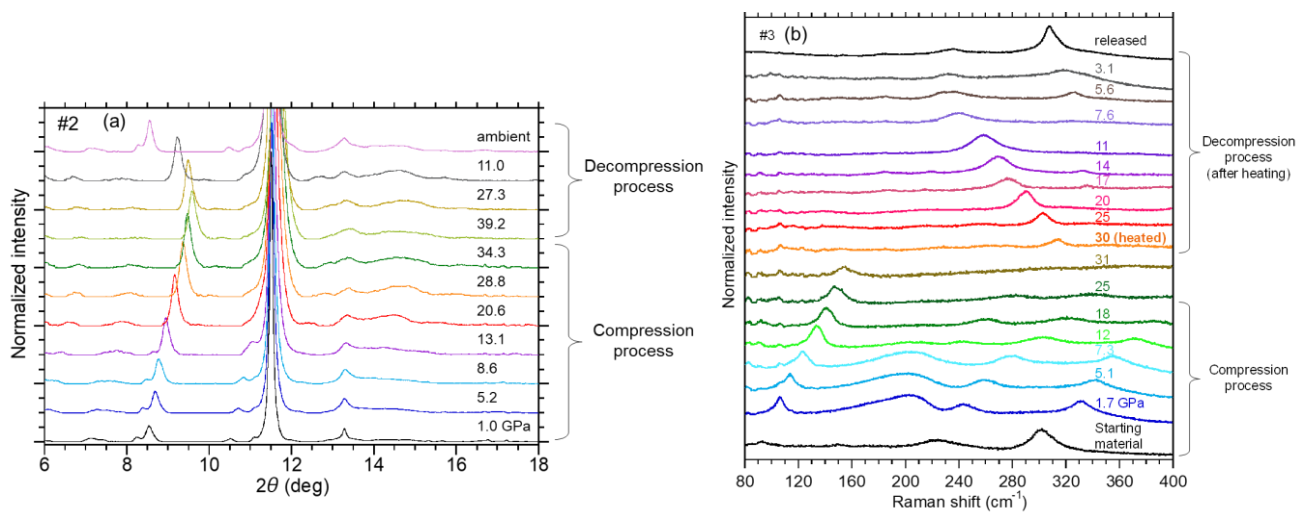

**Figure S2.** (a) XRD patterns of the starting material without heating treatment under various pressures. (b) Raman spectra of starting materials under compression process and decompression process after the synthesis.

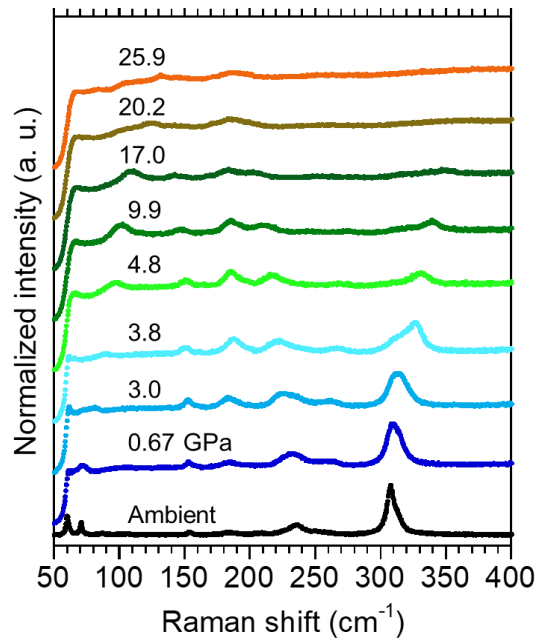

**Figure S3.** Raman spectra of $Sn_2S_3$ under high pressures.